\documentclass[12pt]{article}

\pdfoutput=1

\usepackage{amsmath,amssymb,enumerate,graphicx,color,caption, soul}

\newcommand{\amp}{&\!\!}
\newcommand{\beq}{\begin{equation}}
\newcommand{\eeq}{\end{equation}}
\newcommand{\bea}{\begin{eqnarray}}
\newcommand{\eea}{\end{eqnarray}}
\newcommand{\m}{m_F}
\newcommand{\F}{\psi_F}
\newcommand{\bF}{\bar{\psi}_F}
\newcommand{\Q}{Q_F}
\newcommand{\n}{n_F}
\newcommand{\mpl}{M_{Pl}}
\newcommand{\meson}{\pi_F}
\newcommand{\bmeson}{\bar{\pi}_F}
\newcommand{\B}{B_F}
\newcommand{\bB}{\bar{B}_F}
\newcommand{\betaF}{\beta_{\psi_F}}
\newcommand{\betam}{\beta_{\pi_F}}
\newcommand{\betaB}{\beta_B}
\newcommand{\Qmeson}{Q_{\pi_F}}
\newcommand{\Qatom}{Q_{A_F}}
\newcommand{\Qbatom}{Q_{\bar{A}_F}}
\newcommand{\nGalaxy}{n^{\rm gal}_B}
\newcommand{\nmeson}{n_{\pi_F}}
\newcommand{\zmeson}{\zeta_{\pi_F}}
\newcommand{\nearth}{n_{\pi_F}^{\rm ea}}
\newcommand{\nearthinit}{n_{\pi_F0}^{\rm ea}}
\newcommand{\nstar}{n_{\pi_F}^{\rm star}}

\begin{document}

\title{\bf Astrophysical Constraints on \\ Singlet Scalars at LHC}

\author{
\Large{Mark P.~Hertzberg, Ali Masoumi} \\
~\\
{\em Institute of Cosmology, Department of Physics and Astronomy},\\
{\em Tufts University, Medford, MA 02155, USA}}

\date{\today}

\maketitle

\begin{abstract}
We consider the viability of new heavy gauge singlet scalar particles at colliders such as the LHC. Our original motivation for this study came from the possibility of a new heavy particle of mass $\sim$\,TeV decaying significantly into two photons at colliders, such as LHC, but our analysis applies more broadly. We show that there are significant constraints from astrophysics and cosmology on the simplest UV complete models that incorporate such new particles and its associated collider signal. The simplest and most obvious UV complete model that incorporates  such signals is that it arises from a new singlet scalar (or pseudo-scalar) coupled to a new electrically charged and colored heavy fermion. Here we show that these new fermions (and anti-fermions) would be produced in the early universe, then form new color singlet heavy mesons with light quarks, obtain a non-negligible freeze-out abundance, and remain in kinetic equilibrium until decoupling. These heavy mesons possess interesting phenomenology, dependent on their charge, including forming new bound states with electrons and protons. We show that a significant number of these heavy states would survive for the age of the universe and an appreciable number would eventually be contained within the earth and solar system. We show that this leads to detectable consequences, including the production of highly energetic events from annihilations on earth,  new spectral lines, and, spectacularly, the destabilization of stars. The lack of detection of these consequences rules out such simple UV completions, putting pressure on the viability of such new particles at LHC. To incorporate such a scalar would require either much more complicated UV completions or even further new physics that provides a decay channel for the associated fermion.
\let\thefootnote\relax\footnotetext{Electronic address: {\tt mark.hertzberg@tufts.edu, ali@cosmos.phy.tufts.edu}}
\end{abstract}

\newpage

\section{Introduction}\label{Introduction}
The Standard Model of particle physics, minimally coupled to gravity, is an effective field theory that is an adequate description of most, if not all, phenomena at low energies. However, we know it also has several shortcomings, including the lack of a dark matter candidate, hierarchy problem, vacuum stability, baryogenesis, gauge coupling unification, strong CP problem, problematic UV behavior, etc. Hence we anticipate that there is new physics at high energies, possibly several 100s of GeV or several TeV and beyond. This new physics may resolve some of these shortcomings, especially those that pertain to the weak sector. Within the framework of relativity and quantum mechanics, ``new physics" ordinarily just means new particles of some mass and spin with various types of couplings to the known Standard Model particles. Of course countless ideas abound, including supersymmetric particles, sterile neutrinos, GUT particles, axions, and so on. 

Perhaps the simplest possibility for new physics is to add one or more new heavy gauge singlet scalars $\phi$. Such singlets can couple to the Standard model at the renormalizable level through the Higgs portal $\Delta\mathcal{L}\propto\phi H^\dagger H,\,\phi^2 H^\dagger H$. Moreover, general effective field theory arguments imply that generically it will also couple to other Standard Model particles through various higher dimension operators. Allowed dimension 5 operators include couplings to photons $F$ and gluons $G$ as
\beq
\Delta\mathcal{L}_{eff}={\phi\over \Lambda}\left(c_1 F^2+c_2\,\mbox{Tr}[G^2]+\tilde{c}_1 F\tilde F+\tilde{c}_2\,\mbox{Tr}[G\tilde G]\right),
\eeq
where the first and second terms apply to scalars and the third and fourth terms apply to pseudo-scalars (the tilde on $F,G$ is used to indicate the dual tensor and the tilde on the couplings is used throughout this paper to indicate the pseudo-scalar case). Here $\Lambda$ is the characteristic scale of the new physics.

An immediate consequence of this general effective field theory reasoning is that as long as the scale of new physics $\Lambda$ is not too large and the mass of the new particle $\phi$ is not too large, then effects associated with this new scalar can be observed at colliders such as the LHC. In particular, the $\sim c_2,\tilde{c}_2$ terms allow $\phi$ particles to be produced at the LHC through gluon fusion, while the $\sim c_1,\tilde{c}_1$ terms allow $\phi$ particles to decay into pairs of photons that could be detected.

For example,  a recent tantalizing signal with roughly $3\,\sigma$ local significance at both CMS \cite{CMS} and ATLAS \cite{ATLAS} at LHC is an excess in pairs of photons at an energy of $\sim 750$\,GeV. The most basic interpretation of the signal would be a new scalar (or pseudo-scalar) with mass $m_\phi\sim 750$\,GeV. Many interesting papers on this topic exist in the literature; for example see [3 - 38] and references therein.\footnote{Our paper was released before the signal at $750$\,GeV went away. This is in accord with our results, which indicate that such new particles are difficult to accommodate with astrophysical constraints.}

This provides an initial motivation for this paper, but our analysis is much more general and applies to a range of possible new scalars. Our {\em main} motivation is to identify  the very significant constraints coming from astrophysics and cosmology on the  viability of such new particles.\footnote{Similar considerations were used to constrain axion models \cite{DiLuzio:2016sbl}.}

At first sight it may seem that little could be gleaned from astrophysical observations, because such  particles decay rapidly and therefore would not be  dark matter candidates or play any cosmological role. On the other hand, we know that in order to observe such particles at LHC would require a $\Lambda$ that is not very large, perhaps tens of TeV or so. This would imply that the effective theory used here is being applied at a scale not too far from its cutoff. So it is very important to provide a UV completion of such models. The UV completion means to simply insert even more new particles with various masses and spins that pushes the cutoff to much higher scales. 

In this paper we investigate the consequences of the most obvious and simplest UV completions. These involve the introduction of a new fermion $\F$ with mass $\m=\mathcal{O}$(TeV). The requirement that the branching ratio of $\phi$ decays into photons is large means the fermion should carry a significant electric charge (or hyper-charge from the high energy point of view), and the production from gluon fusion means the fermion should carry color. We investigate the astrophysical implications of this new fermion. In particular, we find that these fermions will be produced thermally in the early universe, then lock-up with light quarks to form new types of heavy mesons. We compute the heavy meson freeze-out abundance and find it to be $\mathcal{O}(10^{-6})$ of the matter density of the universe. These heavy mesons possess interesting phenomenology and can capture protons or electrons depending on the charge. We find that in any case, they maintain kinetic equilibrium with the cosmic plasma and then live for the age of the universe. We show that an appreciable number will be contained within the earth and sun over the course of the formation and history of the solar system. We argue that this leads to significant physical effects, such as annihilations on the earth, new spectral lines, and destabilization of stars, which rules out such models. So the astrophysical constraints are very  significant. Finally, we discuss the implications for model building and the LHC.

\section{Simple Models}\label{SimpleModels}
A two photon resonance at a collider such as the LHC requires a very specific spin for the decaying particle. Obviously it cannot be a fermion by conservation of angular momentum, nor can it be a spin 1 particle by Bose-Einstein statistics, as summarized by the Landau-Yang theorem. The only remaining possibilities are spin 0 and spin 2. Although one could imagine that it is some Kaluza Klein spin 2 mode associated with the graviton \cite{Han:2015cty}, the spin 0 option is the more common scenario and will be the focus of our paper, which we will denote $\phi$.

We also need to enquire as to the particle's renormalizable couplings to the Standard Model, and in particular, whether it carries any gauge charges. At first sight, a decay into two photons is reminiscent of the Higgs particle, which couples directly to $W^{\pm}$ and $Z$ bosons. But the same cannot be true for $\phi$. The reason is that the Higgs particle can be organized into a complex doublet as is appropriate to provide the UV completion of massive spin 1 particles, $W^{\pm}$ and $Z$. Without any further evidence, there is no reason why $\phi$ would be part of some multiplet in the visible sector. Hence we take it to be a gauge singlet with respect to the Standard Model gauge group. It might carry charges in some hidden sector, but for simplicity, let us imagine that it is just a single scalar.

\begin{figure}[t]
\begin{center}
\includegraphics[width=10cm]{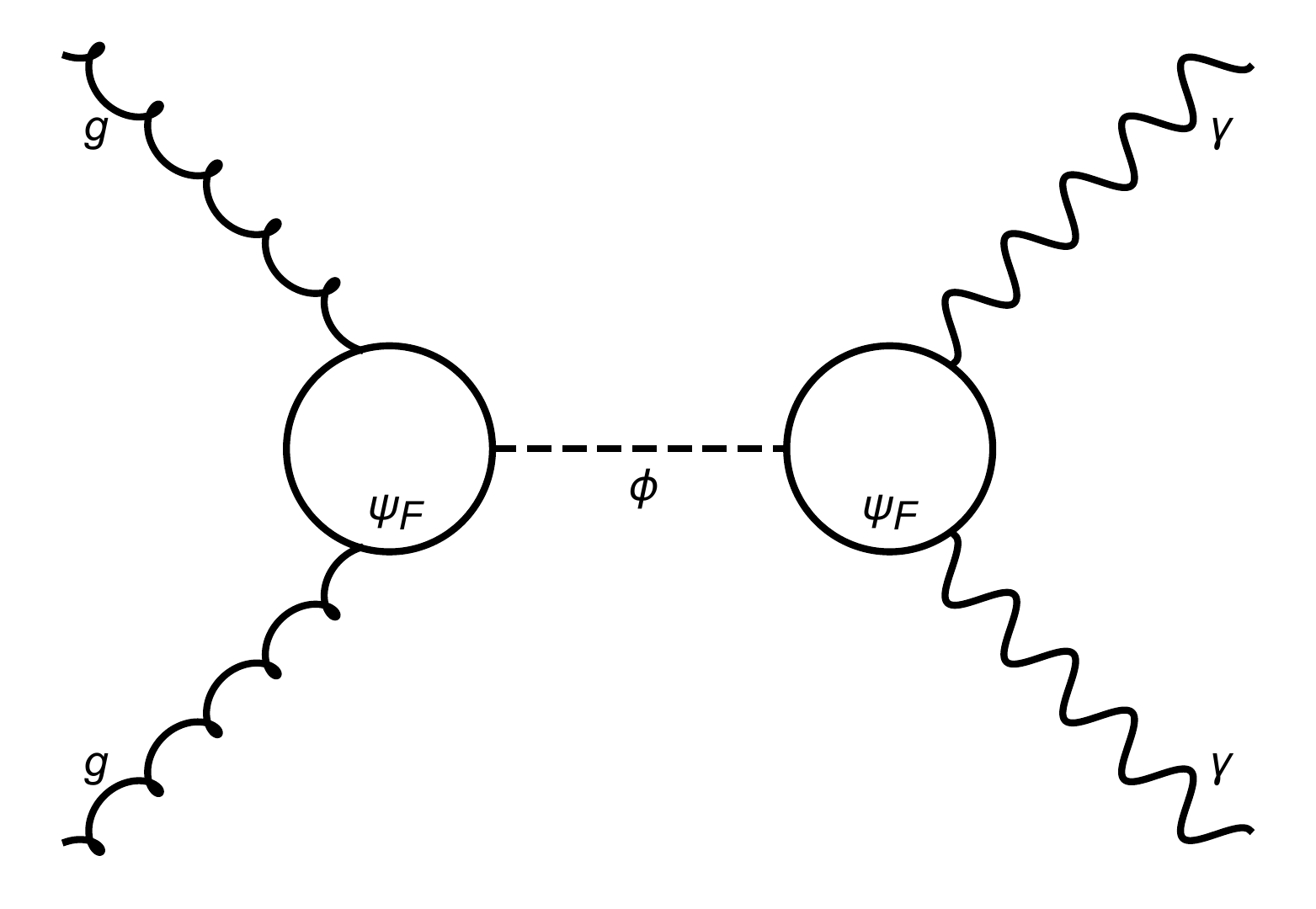}
\end{center}
\caption{Production and decay process $gg\to\phi\to\gamma\gamma$: Gluons $g$ fuse into an off-shell $\phi$ via a fermion loop $\F$, which then decays into photons $\gamma$ via a fermion loop $\F$.}
\label{GluonFusionPhotonDecay}\end{figure}

Renormalizable couplings to the Standard Model gauge bosons must then be provided by some additional new particle/s. The Occam's razor model would be to only add one new particle. This can be a spin 1/2 fermion $\F$, carrying various charges under the Standard Model gauge group. We assume $SU(3)$ color for production from gluon fusion at proton factories like LHC and $U(1)$ electric (hyper) charge allows decay into photons; see Figure \ref{GluonFusionPhotonDecay}. It is possible that it carries weak $SU(2)$ charge too, though we will mainly focus on the singlet case. 

The UV complete model is given by the following
\bea
\mathcal{L}  \amp = \amp \mathcal{L}_{SM}+ {1\over 2}(\partial\phi)^2-{1\over 2}m_\phi^2\phi^2-V_I(\phi,H)\nonumber\\
\amp\amp+i\,\bF\gamma^\mu D_\mu\F-\m\bF\F+\phi(y_F\bF\F+\tilde{y}_F\bF\gamma_5\F),
\label{Lagrangian}\eea
where $\mathcal{L}_{SM}$ is the usual Standard Model Lagrangian. Here $V_I$ is an interaction potential including terms $\propto \phi^3,\phi^4,\phi H^\dagger H,\phi^2 H^\dagger H$. We are assuming that $\phi$ is being expanded around $\phi=0$ and so will not pick up a VEV in this basis, so such terms will not play an important role. Also $D_\mu$ is the covariant derivative which minimally couples $\F$ to $U_Y(1)$ fields with hyper-charge $Y_F$ (electric charge $\Q$) and to gluons with coupling $g_s$. We allow a Yukawa like interaction that couples the new scalar to the new fermion ($y_F$ for scalar and $\tilde{y}_F$ for pseudo-scalar). We note that the fermion should be vector-like to allow for this renormalizable coupling to a gauge singlet. 

The production cross-section of di-photons from a $\phi$ resonance can be made significant by assuming that the Yukawa coupling $y_F,\tilde{y}_F$ is large, since the cross-section scales as 
\beq
\sigma(pp\to\phi\to\gamma\gamma) \propto y_F^2,\tilde{y}_F^2.
\eeq
By an appropriate choice of $y_F,\tilde{y}_F$ we can arrange the production rate of photons to accommodate a di-photon excess at some energy by choosing the corresponding mass $m_\phi$ to equate with the invariant combined energy of the photon pair. In the case of the LHC, an interesting mass range is $m_\phi\sim$\,TeV. This can be achieved with perturbative values of the Yukawa coupling so long as the process is not severely off-shell, so this means the fermion should not be extremely heavy. On the other hand, the fermion cannot be extremely light or we would have directly detected such a colored/charged particle by now. This pushes the mass of the new fermion to be in a window around $\m=\mathcal{O}$(TeV).

In this model, there are obviously multiple decay channels for the $\phi$ particle. It cannot decay into an on-shell pair of $\F$ and $\bF$ as we take $m_\phi<2\,\m$. Nevertheless, it can decay into various Standard Model particles through a fermionic loop. In particular, $\phi$ can decay into pairs of photons, pairs of gluons, one photon and one gluon, pairs of $Z$ bosons (since the fermion carries hyper-charge), one photon and one $Z$ boson, and one gluon and one $Z$ boson. 

In order for the photon resonance to be significant at LHC, we require that the branching ratio into photons be appreciable. At leading order, we can estimate the ratio of decay rates into pairs of photons versus pairs of gluons as \cite{PeskinSchroeder}
\beq
r\equiv{\Gamma(\phi\to\gamma\gamma)\over\Gamma(\phi\to g g)}\approx{N_c^2\,\alpha^2 \,\Q^4\over2\,\alpha_s^2 \,\chi},
\eeq
where $\chi\sim 1.5$ is a fudge factor that accounts for loop-level QCD effects, capturing the corrections from an otherwise tree-level result. For a dominant resonance in the di-photon channel, this ratio should satisfy the bound $r\geq1$. We note that $r$ depends very sensitively on the charge $\Q$ as it depends on it to the fourth power. To satisfy the bound we obtain
\beq
|\Q|\gtrsim2.
\eeq
(Related discussion appears in Ref.~\cite{Pilaftsis:2015ycr}) This bound on the charge is not extremely tight, as it would require a detailed analysis of various QCD backgrounds to determine at what level one should have already seen a di-jet signal, but it is suggestive that the charge of the new fermion is appreciable, and probably larger than that of both up-like and down-like quarks. The specific value of the charge, whether it is integer or fractional charge (we shall assume quantization in 1/3 units) will have important consequences on its cosmology, as we explain in the later sections.

\section{Freeze-Out Abundance}\label{FreezeOutAbundance}
This new fermion $\F$, required for the UV completion of the singlet scalar coupling to the Standard Model gauge bosons, can have interesting cosmological consequences. The late time stability of the $\F$ particles will be discussed in the next section. In this section we discuss the initial behavior of these fermions (and anti-fermions) in the very early universe.

\begin{figure}[t]
\begin{center}
\includegraphics[width=5cm]{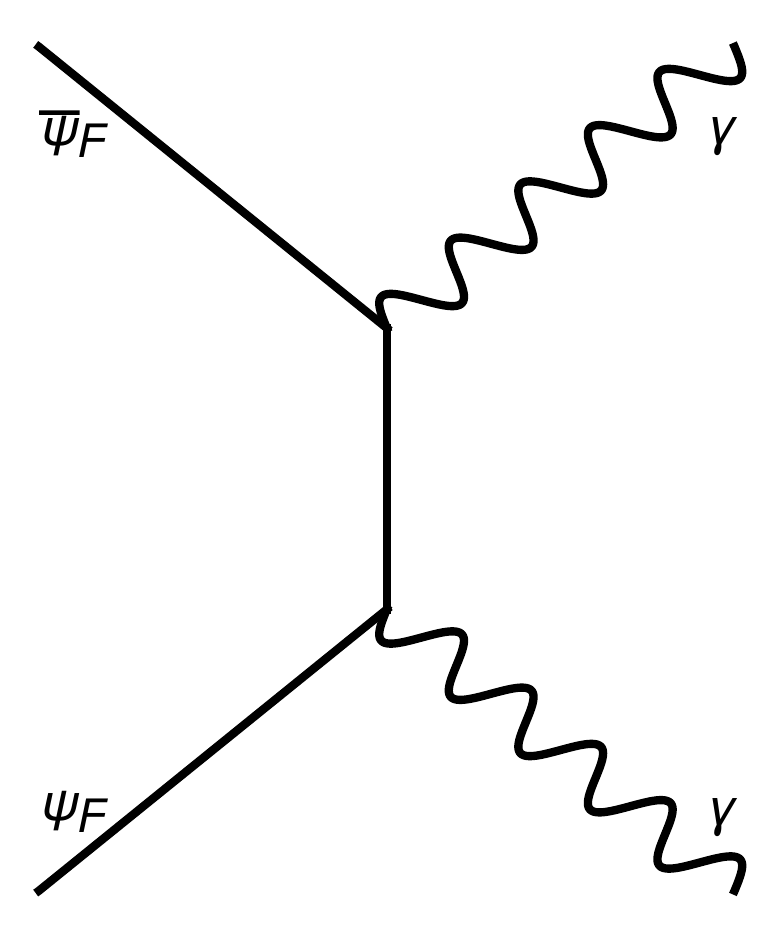}\,\,\,\,\,\,\,\,\,\,\,\,\,\,\,\,\,\,\,\,\,\,\,\,
\includegraphics[width=5cm]{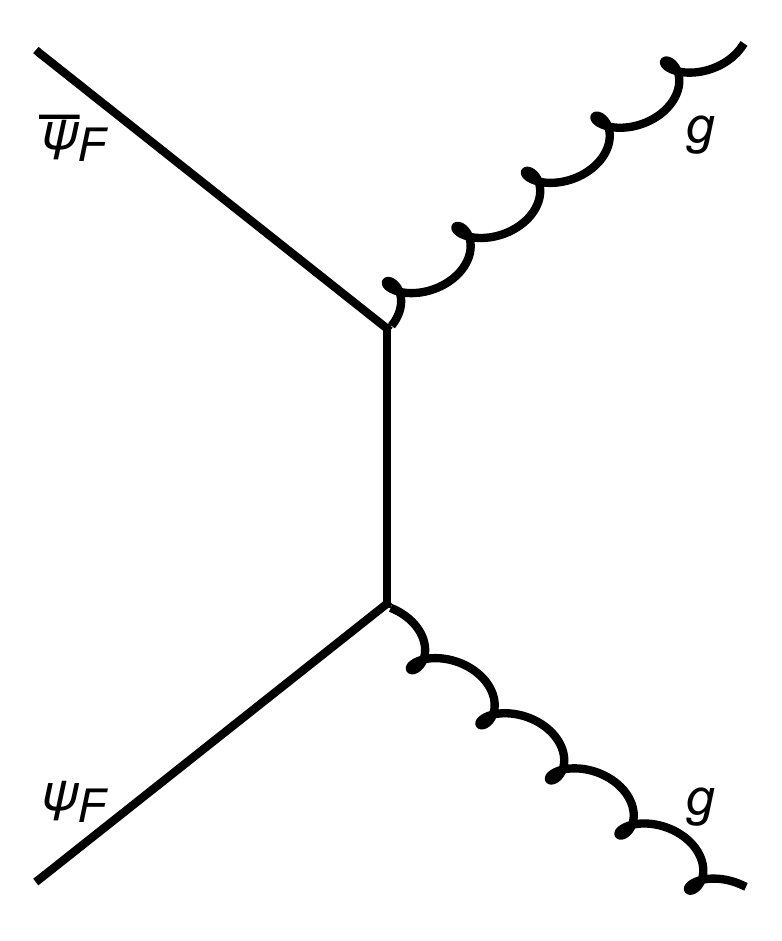}
\end{center}
\caption{Some processes that maintain thermal equilibrium until freeze-out. Left: Fermions annihilate into photons and vice versa $\F \bF\leftrightarrow\gamma\gamma$. Right: Fermions annihilate into gluons and vice versa $\F \bF\leftrightarrow g g$.}
\label{Annihilation}\end{figure}

Since $\F$ carries electric and color charge it will be produced thermally in the early universe and come into thermal equilibrium. Before the QCD phase transition the $\F$ and $\bF$ will be weakly interacting particles, produced by and annihilating into Standard Model particles; see Figure \ref{Annihilation}. For example, the cross-section of annihilation of a fermion/anti-fermion pair into pairs of photons or gluons at temperatures well above the mass of the fermion is 
\beq
\langle\sigma_{ann} |v|\rangle(\F\bF\to\gamma\gamma)\sim {\alpha^2\,\Q^4\over T^2},\,\,\,\,\,\,\langle\sigma_{ann} |v|\rangle(\F\bF\to g g)\sim {\alpha_s^2\over T^2}.
\eeq
In the relativistic regime, the number density is $\n\sim T^3$. This gives an estimate for the annihilation rate as $\Gamma_{ann} =\n\langle\sigma_{ann} |v|\rangle\sim \alpha^2\,\Q^4\,T$ (or $\alpha_s^2\,T$), which is much larger than the Hubble rate in the radiation era which is Planck suppressed $H\sim T^2/\mpl$. This confirms that this species will indeed be initially in thermal equilibrium with the cosmic plasma. 
But as usual for a heavy species, its number density begins to plummet exponentially $\n\sim (\m T)^{3/2}e^{-\m/T}$ once the temperature falls below its mass, leading to a fall out of equilibrium at some temperature $T_f$. 

The freeze-out abundance of $\F$ can be computed using the standard analysis often employed for cold dark matter candidates, by using the Boltzmann equation. If we assume the $\F$ ($\bF$) particles remain after freeze-out (we will return to this issue shortly), then the well known late time abundance formula applies, namely that the freeze-out scales inversely with the annihilation cross-section as 
\beq
\Omega_{\F} h^2 \approx {3\times10^{-38}\mbox{cm}^2\over\langle\sigma_{ann} |v|\rangle_f}{\m\over\sqrt{g_*}\,T_f},
\label{FreezeOut}\eeq
where the freeze-out temperature is typically $T_f\sim\m/20$, the effective number of relativistic species is typically $g_*\sim100$, and $\langle\sigma_{ann} |v|\rangle_f\sim \alpha_s^2/\m^2$ is the annihilation cross-section computed at freeze-out. For $\m\sim$\,TeV and $\alpha_s\sim10^{-1}$ this leads to a relic density that is perhaps an order of magnitude smaller than the observed dark matter density $\Omega_M h^2\approx 0.12$. Let us parameterize this with a quantity $\betaF$ by writing 
\beq
\Omega_{\F}=\betaF\,\Omega_M,\,\,\,\,\mbox{with}\,\,\,\,\betaF=\mathcal{O}(10^{-1}).
\eeq
However, this is a huge over-estimate of the relic abundance, for reasons we now explain.

As the temperature cools and approaches the QCD phase transition $T\sim$\,GeV, the colored $\F$ and $\bF$ particles become strongly interacting. This forces them to lock up into color singlets, as a type of heavy hadron. Several possibilities exist, including (i) $\F$ binding with a light Standard Model anti-quark $\bar{q}$ to form a heavy meson $\meson=\F\,\bar{q}$ (similarly, anti-meson $\bmeson=\bF\,q$), (ii) $\F$ binding with two light Standard model quarks $q_1$, $q_2$ to from a heavy baryon $\B=\F q_1 q_2$ (similarly, anti-baryon $\bB=\bF \bar{q}_1 \bar{q}_2$), (iii) super heavy mesons $\F\bF$, (iv) super heavy baryons $\F\F q$ (similarly, $\bF\bF\bar{q}$), etc. 

The formation of these color singlet states occurs as the temperature drops well below the mass of the fermion. Thermodynamic considerations then strongly favors the formation of the lightest mesons (i) $\meson=\F\,\bar{q}$ and anti-mesons $\bmeson=\bF\,q$, since the formation of super heavy hadrons (involving multiple $\F$'s) is Boltzmann suppressed and the formation of baryons with 3 particles is entropically suppressed compared to the formation of mesons with 2 particles. The weak interaction will generically lead to the Standard Model quark involved in the meson decaying into the first generation quark, presumably the lightest quark $u$. It is conceivable that if $\F$ is negatively charged, the different electrical energy associated with the $d$ quark could render the $\F\,\bar{d}$ meson stable instead, which we will also mention when appropriate.

The QCD effects on the mass of the bound states is around the QCD scale which is of order of GeV. Therefore, these mesons $\meson$ (and anti-mesons $\bmeson$) have a mass that is almost entirely determined by the mass of the new heavy fermion; so $m_{\meson}\approx\m$. Their size is roughly set by the QCD scale: $R_{\meson}=\mathcal{O}(1/$GeV) \cite{Kang:2006yd}. This means that they act as  large objects compared to their Compton wavelength, increasing their probability to find their anti-particles. The associated enhanced annihilation cross-section is estimated as
\beq
\langle\sigma_{ann} |v|\rangle(\meson\bmeson\to \mbox{SM}\,\mbox{SM})\sim 0.03/\mbox{GeV}^2,
\eeq
where we assumed $|v|\sim\sqrt{T_f/\m}\sim\sqrt{1\,\mbox{GeV}/1\,\mbox{TeV}}\approx 0.03$.
This annihilation cross-section of mesons is much larger than the annihilation of free $\F$'s before confinement, which we estimated earlier as $\langle\sigma _{ann}|v|\rangle(\F\bF\to gg)\sim\alpha_s^2/\m^2\sim 10^{-8}/$GeV$^2$. As a rough estimate of the freeze-out meson abundance, we insert this much higher cross-section into the standard freeze-out formula in eq.~(\ref{FreezeOut}) and take $T_f\sim$\,GeV to give
\beq
\Omega_{\meson}=\betam\,\Omega_M,\,\,\,\,\mbox{with}\,\,\,\,\betam=\mathcal{O}(10^{-6}).
\eeq
This is a small, but non-negligible abundance, whose consequences we shall describe.

\section{Cosmic Survival}\label{CosmicSurvival}
It is important to examine the potential for survival of these mesons. According to the Lagrangian given in eq.~(\ref{Lagrangian}), the $\F$ particles, and the associated $\meson$ mesons, appear stable. But the Lagrangian may be incomplete. In particular, one might try to couple this colored fermion directly to quarks by operators such as $\Delta\mathcal{L}\sim P_L\bF\,q_R+h.c.$, assuming for simplicity that $\F$ is an $SU(2)$ singlet, where $P_L$ is the left handed projection operator (other related possibilities arise for an $SU(2)$ doublet $\Delta\mathcal{L}\sim P_L\bF\!\cdot\!H\, q_R+h.c.$). This means $\F$ mixes with the quarks, allowing decays. If the electric charge of $\F$ is $-1/3$ then it can couple to down quarks in this way; if it has charge $+2/3$ it can couple to up quarks in this way. However, as we discussed earlier, in order for the resonance into di-photons to be the first dominant signal seen at a proton factory like the LHC, the charge may need to be larger than this $|\Q|\gtrsim2$. In this case, direct mixing with quarks (or leptons) at the renormalizable level is forbidden (More precisely, it is forbidden for $|\Q|\geq1$ for an $SU(2)$ singlet and for $|Y_F|\ge3/2$ for an $SU(2)$ doublet). This implies that the $\F$ particles, and the associated $\meson$ mesons, are indeed stable. (We will return to this issue in the discussion section.)

These heavy mesons $\meson=\F\,\bar{u}$ carry electric charge
\beq
\Qmeson = \Q-2/3,
\eeq
which is evidently non-zero for $|\Q|\geq1$ (similarly for $\meson=\F\,\bar{d}$). If $\Qmeson>0$, these mesons will capture some number $N_{e^-}$ of electrons (although this might take until recombination for the elections to not be ionized), or if $\Qmeson<0$, these mesons will capture some number $N_p$ of protons (which can happen much earlier than recombination). Since there are both mesons and anti-mesons, one of them will capture electrons while the other will capture protons. The binding energy of a proton to the negatively charged meson is $\Delta E_F\approx{1\over2}m_p\Q^2\alpha^2$, which is much larger than the binding energy energy of hydrogen $\Delta E_H={1\over2}m_e\alpha^2$. So the capture of the proton is essentially guaranteed to occur in the early universe (and by conservation of charge the electron will be eventually captured by the oppositely charge meson too).

We shall refer to these as `dressed' mesons or `atoms'. If $\Qmeson$ is integer, then the resulting heavy atom can become neutral after capture. If $\Qmeson$ is non-integer, then the resulting heavy atom will remain charged. Let's call the total charge (which can change before and after recombination)
\beq
\Qatom=\Qmeson+N_p-N_{e-}.
\eeq
If this vanishes, then these neutral heavy atoms will inevitably live for a long time without meeting their anti-particles. However, if the net charge is non-zero, one might be concerned that this electrical interaction will cause a (dressed) meson $\meson=\F\,\bar{u}$ to meet a (dressed) anti-meson $\bmeson=\bF\,u$ and annihilate. (Or a similar story if $\meson=\F\,\bar{d}$ is the stable meson). However this is highly unlikely, as we now explain.

\begin{figure}[t]
\begin{center}
\includegraphics[width=5cm]{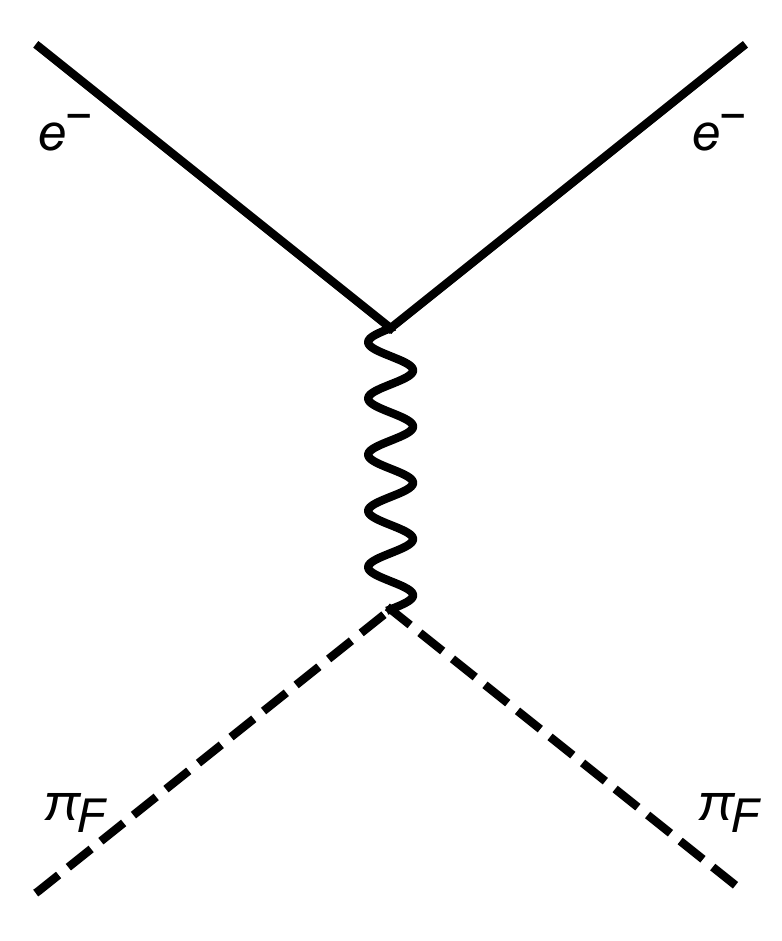}\,\,\,\,\,\,\,\,\,\,\,\,\,\,\,\,\,\,\,\,\,\,\,\,
\includegraphics[width=5cm]{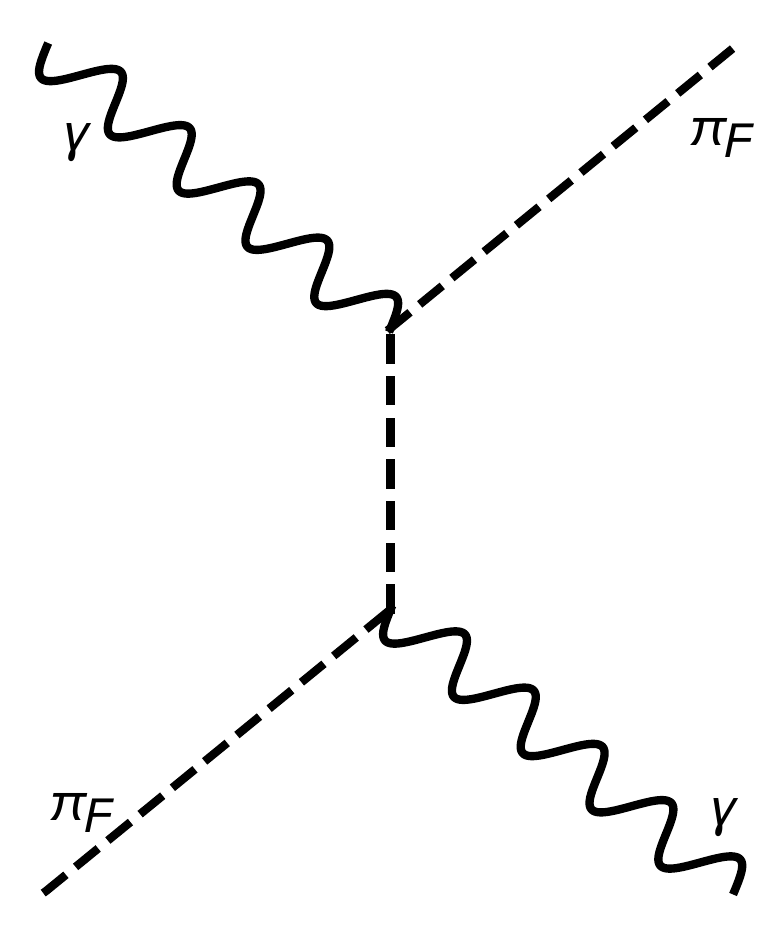}
\end{center}
\caption{Some processes that couple mesons to the cosmic plasma. Left: Mesons interact with electrons via Rutherford scattering $\meson\,e^{-}\leftrightarrow \meson\,e^{-}$. Right: Mesons interact with photons via Thomson scattering $\meson\,\gamma\leftrightarrow \meson\,\gamma$.}
\label{Scattering}\end{figure}

For the charged (dressed) mesons, although the annihilations have frozen out, scattering off the cosmic plasma can still be significant. The mesons can interact with cosmic electrons (and protons) via Rutherford scattering (see left hand side of Figure \ref{Scattering}) and cosmic photons by Thomson scattering (see right hand side of Figure \ref{Scattering}). 
The Rutherford scattering cross-section $\sigma_{R}(\meson\,e^-\to\meson\,e^-)$ is relatively large, though the number of electrons is relatively small. While the Thomson scattering cross-section $\sigma_{T}(\meson\,\gamma\to\meson\,\gamma)$ is relatively small, though the number of photons is relatively large. Since mesons are heavy, the $\meson+e^-$ scattering is controlled by the electron energy $E_{e^-}$. The differential cross-section is given by
\beq
{d\sigma_R\over d\Omega}(\meson\,e^-\to\meson\,e^-)={\alpha^2\,\Qatom^2\over 4\,E_{e^-}^2}{\cos^2(\theta/2)\over\sin^4(\theta/2))}.
\eeq
For late times $T<m_{e}$, but before recombination $T> T_{rec}\approx 0.3$\,eV, the electrons are free and non-relativistic. The energy $E_{e^-}$ is of the order the photon temperature $T$, by equipartition, since the electrons maintain thermal equilibrium in the cosmic plasma. After $e^+\,e^-$ annihilation, the number density of electrons can be estimated by the baryon-to-photon ratio $\eta\approx6\times 10^{-10}$ as
\beq
n_{e^-}\sim\eta\, n_\gamma\sim\eta\,T^3.
\eeq
The typical relative velocity of thermal electrons is $|v|\sim\sqrt{T/m_e}$. Using $\Gamma_R=n_{e^-}\langle\sigma_R |v|\rangle$, this gives a Rutherford scattering rate of mesons as
\beq
\Gamma_R(\meson\,e^-\to\meson\,e^-) \sim \eta\,\alpha^2\,\Qatom^2\sqrt{T^3\over m_e}.
\eeq
Using $m_e\sim 0.5$\,MeV, $\alpha\sim 1/137$, this gives $\Gamma_R\sim 10^{-17}\Qatom^2\sqrt{T^3/\mbox{eV}}$. In the radiation era the Hubble parameter is $H\sim T^2/\mpl\sim 10^{-27}\,T^2/\mbox{eV}$. Now recall that matter-radiation equality is at a temperature $T_{eq}\sim 1$\,eV. So for non-zero charge $\Qatom$, we have $\Gamma_R\gg H$ in this era. It is simple to see that this continues to hold up to recombination $T_{rec}\sim 0.3$\,eV. 

So (dressed) charged mesons maintain kinetic equilibrium with the cosmic plasma, at least up to recombination. By equipartition we can determine the typical meson kinetic energy as $K_{\meson}\sim T$. On the other hand, we can estimate the typical nearest neighbor  Coulomb energy between a (dressed) meson of charge $\Qatom$ and a (dressed) anti-meson of charge $\Qbatom$ as 
\beq
V_{\meson}\sim{\alpha\,\Qatom\Qbatom \over r},
\eeq
which is typically attractive (with $\Qatom$ and $\Qbatom$ usually carrying opposite signs). The (inverse) nearest neighbor distance is roughly 
\beq
r^{-1}\sim (n_{\meson})^{1/3}=\left(\betam\rho_M\over\m\right)^{1/3}\!\sim\left(m_p\betam\eta\over\m\betaB\right)^{1/3}\!T,
\eeq
where in the last step we used that the matter density of the universe can be parameterized by CMB temperature as follows: 
$\rho_M=\rho_B/\beta_B\sim m_p\,n_B/\betaB=m_p\,\eta\,n_\gamma/\betaB\sim m_p\,\eta\,T^3/\betaB$, with $\betaB\approx0.2$ is the baryonic fraction of matter. So the ratio of Coulomb potential energy to kinetic energy is roughly time independent with value
\beq
{V_{\meson}\over K_{\meson}}\sim \alpha\,\Qatom\Qbatom\!\left(m_p\betam\eta\over\m\betaB\right)^{1/3}.
\eeq
For typical values, this gives $V_{\meson}/K_{\meson}=\mathcal{O}(10^{-8})$. Hence the mesons carry far too much kinetic energy to be captured by the Coulomb energy of their anti-meson neighbor; thus avoiding annihilation before recombination. 

One can check that after recombination, until present day, these heavy charged atoms continue to avoid capture and annihilation. In fact Thomson scattering off CMB photons continues to inject kinetic energy into the mesons, even though Rutherford scattering off electrons has halted. Of course the neutral heavy atoms easily survive also.

\section{Detectable Consequences}

In this section we mention several physical consequences that these heavy atoms can have in the late universe. Although they comprise only $\betam=\mathcal{O}(10^{-6})$ of the mass density of the universe, which equates to a very small relative number density of heavy atoms to baryons of
\beq
\zmeson\equiv {\nmeson\over n_B}=\mathcal{O}(10^{-8}),
\eeq
(using $\m/m_p=\mathcal{O}(1000)$ and $\rho_B/\rho_M=\mathcal{O}(10^{-1})$) we will see that they still have important observable consequences.

\subsection{Annihilations on Earth}
As these heavy atoms move throughout the galaxy, some number will hit the earth's atmosphere as a type of exotic cosmic ray. Since these heavy atoms have electric interactions, they scatter off molecules in the atmosphere. The scattering cross-section depends sensitively on whether the heavy atom is a heavy $\meson$ (or $\bmeson$) that captured electrons $N_{e^-}$, or its anti-particle $\bmeson$ (or $\meson$) counterpart that only captured protons $N_p$. As a lower bound, let's focus on scattering off neutral molecules. Then we can estimate $\sigma_{scatt}\sim R^2$ the effective size of the smaller of the 2 particles in the collision. If the heavy atom is charged or if the meson has captured an electron, we can estimate it as $R_e\sim(\alpha_e m_e)^{-1}\sim 10^{-10} { \rm m}$, or in the special case in which the heavy atom is neutral and the meson has captured a proton we can estimate it as $R_p\sim(\alpha_e m_p)^{-1}\sim 10^{-13} { \rm m}$. 

The mean free path is $\ell=(n_{\rm atm}\, \sigma_{scatt})^{-1}$, with $n_{\rm atm}\sim 10^{26} {\rm m}^{-3}$ is the number density of gas molecules in the atmosphere. This gives $l\sim 10^{-6}$\,m for $R_e$ and $l\sim 1$\,m for the special case $R_p$. This means that as soon as $\meson$ touches the earth atmosphere it thermalizes with the air molecules and loses its kinetic energy. Because these are much heavier than the air molecules they will deposit near the earth surface. The number density of these particles in thermal equilibrium near the earth surface is given by the Boltzmann formula
\begin{equation}
	\nearth(h) \propto e^{-\m g\, h/T},
\end{equation}
Therefore they are confined in  a height $h_{\rm conf} \sim T/(\m g) \sim 300$\,m, using $\m\sim$\,TeV. So any particle that hits the atmosphere would be deposited and stored near the earth surface (the special case in which the heavy atom is a meson orbited by a proton is such a small atom that it may eventually sink through the earth's surface). 

Since there will be both heavy mesons and anti-mesons accumulating within the earth they will begin to annihilate. If the mesons that have captured begin to sink beneath the surface it may lower the rate of annihilation, but we will just take these number densities as estimates of this effect. The total rate of annihilation between two of these particles per unit volume is given by 
\begin{equation}
	\frac{\Gamma_{ann}^{\rm ea}}{\rm V}= (\nearth)^2  \langle\sigma_{ann} |v|\rangle, 
\end{equation}
where $\sigma_{ann}\sim 1/\mbox{GeV}^2$ is the annihilation cross-section and $|v|\sim\sqrt{T/\m}\sim 40$\,m/s.

Note that this is quadratically sensitive to the number density $\nearth$ of heavy mesons and anti-mesons in the earth. This is non-trivial to compute accurately. But we will estimate this from two different mechanism as follows:

\begin{enumerate}[(I)]
\item {\em Accumulation of random heavy atoms moving throughout the galactic halo}. We note that for charged particles, some fraction $f$ of galactic particles can be bent away from the solar system by the combined earth and sun's magnetic field. For solar wind particles, this is estimated to be $f=\mathcal{O}(10^{-3})$, due to the strong fields between the earth and sun \cite{KenLang}. However, on the other side of the earth, away from the sun, it is much less protected by the combined magnetic fields and particles can easily enter, with $f$ much closer to 1. We take $f=\mathcal{O}(10^{-1})$ as a conservative value.
As the earth  moves in its orbit around the sun, it hits the interstellar  $\meson$'s that entered the solar system. As soon as these particles hit the earth's atmosphere, due to the short mean free path in the atmosphere, they thermalize and deposit near the surface of the earth. We do the calculation of decay rates in two different regimes: the annihilation is very slow and the particles deposit and accumulate over time and annihilation is so fast that the rate of entering the earth is the same as rate of annihilation of the particles. In the first scenario the  number density deposited in the layer of thickness $h_{\rm conf}$ after a time $t$ is readily shown to be
\beq
\nearth\sim \frac{v\, t}{ h_{\rm conf}}\,f\,\zmeson\,\nGalaxy \sim 10^{16}\,f\, {\rm m}^{-3},
\eeq
where we took the velocity to be set by the virial speed $v\sim 2\times 10^5$\,m/s, height $h_{\rm conf}\sim 300$\,m, time $t\sim 1$\,Gyr, the local galactic density of baryons to be $\nGalaxy \sim 6\times 10^4$\,m$^{-3}$ (from $\rho^{\rm gal}_M\sim 0.3$\,GeV/cm$^3$), and the fractional number density of heavy mesons to be $\zmeson\sim 10^{-8}$. This gives an annihilation rate
\beq
{\Gamma_{ann}^{\rm ea}\over V}\sim 10^6\,f^2\,  {\rm m}^{-3}\, {\rm s}^{-1}
\label{est1}\eeq
In the second regime, we expect all the annihilation to happen near the earth's surface. Therefore we should report the rate as per area of the earth surface. The total rate of capture by the earth which is the same as the annihilation rate would be $\pi R^2 v\,f\,\zmeson\,\nGalaxy$ and the rate per area is
\begin{equation}
	\frac{\Gamma_{ann}^{\rm ea}}{A}= \frac14v\,f\,\zmeson\,\nGalaxy\sim 10^{-7} f \, m^{-2}s^{-1}~.
\end{equation}
This means that for example in each year per square meter of detectors we see $f$ decay of 1\,TeV particles which should be detectable in experiments like IceCube.

\item {\em Accumulation of atoms as part of solar system formation}. We can think of (I) as essentially a {\em lower} bound on the number density that accumulates on earth. But it is very possible that there would be a much larger number that arise on earth as simply set by the fractional galactic density; the heavy atoms will act roughly similarly to ordinary atoms and simply form part of stars and planets with a relative number density given directly by the background density $\zmeson$. This implies that the initial densities 
\beq
\nearthinit\sim \zmeson\,n_{\rm B}^{\rm ea} \sim 10^{22}\,{\rm m}^{-3},
\eeq
where $n_{\rm B}^{\rm ea}\sim 10^{30}\,{\rm m}^{-3}$ is the number density of baryons in the earth. This number density decays over time as 
\beq
	\nearth = \frac{\nearthinit}{1+ \nearthinit \langle \sigma v\rangle t}~,
\eeq
and the current density would be $\nearth\sim 10^{11}m^{-3}$. This leads to a decay rate 
\beq
	{\Gamma_{ann}^{\rm ea}\over V} \sim 10^{-5} m^{-3} s^{-1},
	\label{est2prime}
\eeq
or in other words one decay per day per cubic meter of the earth which should be easily detectable.
\end{enumerate}
The lower estimate of the annihilation rate of (I) in eq.~(\ref{est1}) is already large and would have been detected by now, as it leads to explosion production of $\gamma$ rays and other Standard Model particles of energy $\sim\m=\mathcal{O}$(TeV). While the higher estimate of (II) in eq.~(\ref{est2prime}) is obviously even much more dramatic.

\subsection{Corrections to Spectra}
The presence of such heavy particles will also collect in various materials on the earth, such as sea water. By carefully analyzing the spectra of  large vats of water, one should easily have detected such impurities by now. The simplest manifestation of these new ``atoms" would be heavy element with masses of $\mathcal{O}$(TeV) with abundance of around one in a billion which we have not observed in nature. The heaviest atoms known to us are unstable atoms of masses around 300\,GeV and we have not observed any chemical component with such high mass. Because these elements get ionized at the same energies as hydrogen, since they essentially have the same binding energy, we would have seen them in mass spectrometry as very heavy ions. This method is especially useful as it fills potential loop-holes in the annihilation arguments of the previous subsection. Namely, it avoids the possible scenario of the proton captured heavy atoms moving to center of earth and separating from the anti-mesons. And it avoids the possibility of the magnetic belt of the sun is better at keeping out one type of charged atom versus the oppositely charged atom. 

Another possibility is that the spectra of astrophysical gas clouds will be slightly altered by the presence of these new heavy particles. Let's for example look at cases where $\meson$ binds to an electron. This would have the same spectrum and chemistry as the hydrogen atom with one difference that the central mass is bigger than a proton. The energy levels of hydrogen-like atoms are given in terms of the meson charge and reduced mass by 
\beq
E_{\rm n}\approx -{\Qmeson^2\alpha^2 \over 2\,n^2} \frac{m_e M}{m_e+M}.
\eeq
For $\Qmeson\neq 1$ this spectrum is clearly separated from the usual hydrogen spectrum. For unit charge $\Qmeson=1$, it is much closer to the usual hydrogen spectrum. However, note the electron  reduced mass in hydrogen, deuterium, and the heavy mesons is given by $ \approx m_e (1-m_e/m_p)$, $m_e(1-m_e/2m_p)$, and $m_e$, respectively. This adds a spectral line with deviation from hydrogen twice the ones from deuterium. In the case when $\meson$ binds to a proton, the binding energy is
\beq
E_n\approx-{\Qmeson^2\alpha^2\over 2\,n^2}m_p,
\eeq
which is a very clear new set of spectral lines for any $\Qmeson$.
However, it may prove to be too difficult to see this particular effect in the sky as the abundance $\zmeson\sim10^{-8}$ may be too small. But if in some places these particles have high concentration, for example if there are starts of these types, there may be some slight chance of observing these new spectral lines.

\subsection{Destabilization of Stars}

\begin{figure}[htbp] 
   \centering
   \includegraphics[width=4in]{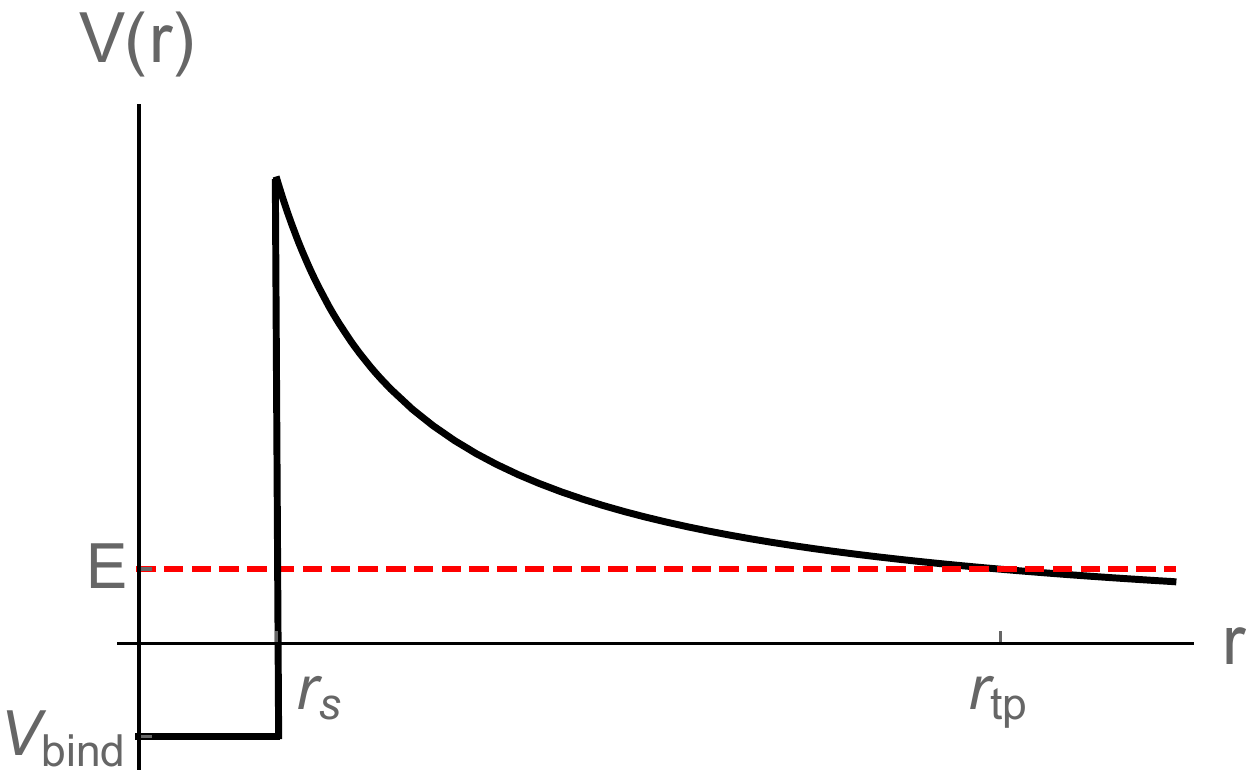} 
   \caption{Potential between two charged nuclei that prevents instantaneous fusion. The strong force acts on a short range $r \le r_s$ and outside of this region the electrostatic repulsion of the two nuclear materials creates a potential barrier that extends to the classical turning point $r_{\rm tp}$. For kinetic energies at the center of the stars $r_{\rm tp}\gg r_s$. The depth of the well on the left side is given by the strong binding energy.}
   \label{fig:potential}
\end{figure}

Just as interesting consequences can occur inside the earth and in astronomical gas clouds, here we discuss the implication for fusion inside stars. Firstly, recall that fusion of protons inside stars is somewhat slow due to the potential barrier provided by the electric repulsion of the protons. The first step in the fusion cycle is $p+p\to D+e^{+}+\nu$, where $D=n\,p$ is deuterium. The electric energy between the two protons at separation $r$ is $V_{pp}=\alpha/r$ and it extends to a radius $r_s \approx 10^{-15}m = 1 {\rm fm}$ where the strong force becomes dominant. The simplest model for this potential is shown in Fig.\ref{fig:potential}. The electric potential energy at $r_s$ is $V_{pp}= \alpha \times {\rm 1\,GeV} \approx 10 {\rm MeV}$. Comparing this to the kinetic energy at the center of the sun with a temperature of roughly 1\,keV, it is clear that the chance of thermal crossing at the tail of the Maxwell-Boltzmann distribution which is roughly ${\cal O}(10^{-1000})$ is negligible. What makes the fusion possible is quantum tunneling through this barrier. To make the comparison easier, let's make an estimate of this tunneling rate for the case that a particle of kinetic energy $E$ and charge $Q$ crosses the barrier. In this case the electric repulsion creates a potential barrier given by $V_{b}= \alpha Q/r$. The WKB approximation of the rate is given by
\begin{equation}\label{WKB}
	P \approx \exp\left({-2 \int_{r_{\rm tp}}^{r_s}}\sqrt{2 \mu [V_b(r)-E]}dr\right)\approx e^{-\pi r_{\rm tp}\sqrt{2\mu E}} = e^{-\pi \alpha Q \sqrt{2\mu /E}}~.
\end{equation}
Here $\mu$ is the reduced mass of the system of the two particles and we also used the approximation $r_s \ll r_{\rm tp}$. 
Now consider the case in which some number of the heavy mesons is inside the star. We know that either the meson (or anti-meson) is negatively charged $\Qmeson<0$ and will have a captured a tightly bound proton in the early universe to form a heavy atom of charge $\Qatom=\Qmeson+1<1$.  This means that the electric energy between this heavy atom and a free proton in the star is 
\beq
V_{A_F p}={\Qatom\alpha\over r}<V_{pp},
\eeq 
(in fact it could even be zero or negative in which there is no barrier to prevent the instantaneous fusion). This provides a significant reduction in the electric potential barrier between the proton captured by the meson and the free proton in the star. Since the rate of fusion is exponentially suppressed by any potential barrier, this alleviates the suppression by a huge amount compared to the usual fusion of 2 free protons. From \eqref{WKB} we can calculate the enhancement. Notice  that the reduced mass of the system of $\Qatom$-proton system is very close to $m_p$ and the reduced mass of proton-proton system is $m_p/2$ . 
\begin{equation}
	\frac{\Gamma_{p\Qatom}}{\Gamma_{pp}} = e^{\pi \alpha\left(1- {\Qatom}\sqrt{2}\right) \sqrt{m_{p} /E}}
\end{equation}
Using the energies at the center of the stars which is of order 1\,keV the enhancement for $\Qatom=0$ is $10^{10}$, for $\Qatom=1/3$ is $10^6$, for $\Qatom=2/3$ is  5 and it would be instantaneous for other cases. This is the enhancement in the first reaction in fusion chain. The second reaction in the fusion chain is  $p+D\to ^3\!\!{\rm He}$. The the quantum tunneling for this reaction for  $\Qatom=0,1/3$ and $2/3$ is enhances by $10^{11}$, $10^6$ and $10^2$ and for other values of $\Qatom$ is spontaneous. Similarly the third reactions in the chain is  $^3{\rm He}+ ^3\!{\rm He}\to \!\!^4{\rm He}+2 p$ for which the quantum tunneling of $\Qatom=0,1/3$ is enhanced by $10^{16}$ and $10^3$, strongly suppressed for $2/3$ and for the other values of $\Qatom$ is spontaneous.
The deuteron binding energy $E_D\approx 2\,$MeV is much larger than the binding energy of the deuteron to the meson, which is $E_{A_F}\approx{1\over2}\Qmeson^2\alpha^2 m_D\approx 0.04\,\Qmeson^2$\,MeV. This means the deuteron then escapes the meson due to the huge kinetic energy released in the fusion. This momentarily leaves a free meson, which can then quickly capture another proton, and the process will repeat.

Of course the number density of these heavy atoms inside stars is very small (an upper estimate would be just to re-scale by the cosmic fractional density $\nstar=\mathcal{O}(\zmeson\,n_B^{\rm star})=\mathcal{O}(10^{-8}\,n_B^{\rm star})$, as in (II) for the earth, or perhaps a few orders of magnitude smaller, as in (I) for the earth). But this exponential enhancement of the fusion rates for all cases except for $\Qatom=2/3$  is  so potent that it would still cause the stars to burn through their fuel very rapidly. This would cause the destabilization of stars into the rapid formation of supernovae, etc, throughout the universe.

\section{Discussion}\label{Discussion}
We have seen that the simplest UV completions of new heavy gauge singlet scalars that may be accessible at LHC lead to very rich phenomenology. In particular, the fermion required for the UV completion to couple the scalar to photons and gluons through loops has a complex cosmology. For multiple reasons, including the accumulation on the earth leading to significant annihilations, the alteration of spectra, and the spectacular destabilization of stars, these models are ruled out observationally.

In terms of model building, the immediate way around these problems is to give the fermion a decay channel. To do so we could imagine relaxing the condition $|\Q|\gtrsim2$. This was motivated by providing a dominant di-photon signal, as it is proportional to the fourth power of $\Q$. It is conceivable that $\Q$ could be smaller ($|\Q|=1/3,\,2/3$) and then a renormalizable mixing with up or down quarks can be possible (and other possibilities if the fermion is an $SU(2)$ doublet). This would easily allow rapid decays in the early universe. However, this would require an explanation as to why the $\phi$ particle is not first seen in some other channel, such as di-jets.

If we continue to focus on this high charge scenario, then from the effective field theory point of view, decays of $\F$ are still possible through the introduction of high dimension operators that couple the new fermion to quarks. This is very sensitive to the specific charge of the fermion $\F$. For example if $\F$ has charge $\Q=-4/3$, then we could introduce dimension 6 operators such as $\Delta\mathcal{L}\sim{1\over M^2}\bar{d}\,\F\, \bar{d}\,u+h.c$. This allows $\F$ to decay into 3 quarks with a rate suppressed by the scale of new physics $M$ as
\beq
\Gamma(\F\to d\,d\,\bar{u})\sim {\m^5\over M^4}. 
\eeq
If we take $M$ is be associated with the GUT scale $M\sim 10^{16}$\,GeV, and $\m\sim$\,TeV, this gives the lifetime $\tau=1/\Gamma\sim 10^8$\,Gyr, which is 7 orders of magnitude longer than the age of the universe. So this would not alleviate the problem. On the other hand, by lowering $M$ by a few orders of magnitude, the $\F$ particles, and their associated heavy mesons, would simply decay in the early universe, avoiding all these astrophysical problems. This implies then that even further new physics must come in well below the GUT scale in order to have a compatible model. 

Of course, once this other new physics with its associated set of additional particles is introduced, one must enquire as to its viability with regards to astrophysics, cosmology, and precision collider physics. So while our findings correctly predicted that the signal at 750\,GeV would go away, there are many other complex interesting possibilities to be explored.

\section*{Acknowledgments}
We would like to thank Norman Christ, Jaume Garriga, Gary Goldstein, Alan Guth, Ken Lang, Krzysztof Sliwa, Jesse Thaler, Alex Vilenkin, Erick Weinberg, and especially Ken Olum for helpful discussions. We would like to thank the Tufts Institute of Cosmology for support. AM is supported by a grant from National Science Foundation PHY-1518742.


\begin{thebibliography}{1}



\bibitem{CMS} 
J. Olsen, 
``CMS physics results from Run 2 presented on Dec. 15th, 2015,"
https://indico.cern.ch/event/442432/;
CMS Collaboration, CMS PAS EXO-15-004,
https://cds.cern.ch/record/2114808/files/EXO-15-004-pas.pdf.

\bibitem{ATLAS}
 M. Kado, 
 ``ATLAS physics results from Run 2 presented on Dec. 15th, 2015,"
https://indico.cern.ch/event/442432/;
ATLAS Collaboration, ATLAS-CONF-2015-081,
https://atlas.web.cern.ch/Atlas/GROUPS/PHYSICS/CONFNOTES/ATLAS-CONF-2015-081/.




\bibitem{Pilaftsis:2015ycr} 
  A.~Pilaftsis,
  ``Diphoton Signatures from Heavy Axion Decays at the CERN Large Hadron Collider,''
  Phys.\ Rev.\ D {\bf 93}, no. 1, 015017 (2016)
  [arXiv:1512.04931 [hep-ph]].

\bibitem{Ellis:2015oso} 
  J.~Ellis, S.~A.~R.~Ellis, J.~Quevillon, V.~Sanz and T.~You,
  ``On the Interpretation of a Possible $\sim 750$ GeV Particle Decaying into $\gamma \gamma$,''
  JHEP {\bf 1603}, 176 (2016)
  [arXiv:1512.05327 [hep-ph]].

\bibitem{Backovic:2015fnp} 
  M.~Backovic, A.~Mariotti and D.~Redigolo,
  ``Di-photon excess illuminates Dark Matter,''
  JHEP {\bf 1603}, 157 (2016)
  [arXiv:1512.04917 [hep-ph]].  
  
\bibitem{Bellazzini:2015nxw} 
  B.~Bellazzini, R.~Franceschini, F.~Sala and J.~Serra,
  ``Goldstones in Diphotons,''
  JHEP {\bf 1604}, 072 (2016)
  [arXiv:1512.05330 [hep-ph]].  
  
  \bibitem{Gupta:2015zzs} 
  R.~S.~Gupta, S.~JŠger, Y.~Kats, G.~Perez and E.~Stamou,
  ``Interpreting a 750 GeV Diphoton Resonance,''
  arXiv:1512.05332 [hep-ph].
  
  \bibitem{Martinez:2015kmn} 
  R.~Martinez, F.~Ochoa and C.~F.~Sierra,
  ``Diphoton decay for a $750$ GeV scalar boson in an $U(1)'$ model,''
  arXiv:1512.05617 [hep-ph].
  
  \bibitem{Demidov:2015zqn} 
  S.~V.~Demidov and D.~S.~Gorbunov,
  ``On the sgoldstino interpretation of the diphoton excess,''
  JETP Lett.\  {\bf 103}, no. 4, 219 (2016)
  [arXiv:1512.05723 [hep-ph]].
  
  \bibitem{Fichet:2015vvy} 
  S.~Fichet, G.~von Gersdorff and C.~Royon,
  ``Scattering light by light at 750 GeV at the LHC,''
  Phys.\ Rev.\ D {\bf 93}, no. 7, 075031 (2016)
  [arXiv:1512.05751 [hep-ph]].
  
  \bibitem{Curtin:2015jcv} 
  D.~Curtin and C.~B.~Verhaaren,
  ``Quirky Explanations for the Diphoton Excess,''
  Phys.\ Rev.\ D {\bf 93}, no. 5, 055011 (2016)
  [arXiv:1512.05753 [hep-ph]].
  
  \bibitem{Csaki:2015vek} 
  C.~Csaki, J.~Hubisz and J.~Terning,
  ``Minimal model of a diphoton resonance: Production without gluon couplings,''
  Phys.\ Rev.\ D {\bf 93}, no. 3, 035002 (2016)
  [arXiv:1512.05776 [hep-ph]].
  
  \bibitem{Falkowski:2015swt} 
  A.~Falkowski, O.~Slone and T.~Volansky,
  ``Phenomenology of a 750 GeV Singlet,''
  JHEP {\bf 1602}, 152 (2016)
  [arXiv:1512.05777 [hep-ph]].
  
  \bibitem{Alves:2015jgx} 
  A.~Alves, A.~G.~Dias and K.~Sinha,
  ``The 750 GeV $S$-cion: Where else should we look for it?,''
  Phys.\ Lett.\ B {\bf 757}, 39 (2016)
  [arXiv:1512.06091 [hep-ph]].
  
  \bibitem{Han:2015cty} 
  C.~Han, H.~M.~Lee, M.~Park and V.~Sanz,
  ``The diphoton resonance as a gravity mediator of dark matter,''
  Phys.\ Lett.\ B {\bf 755}, 371 (2016)
  [arXiv:1512.06376 [hep-ph]].
  
  \bibitem{Cho:2015nxy} 
  W.~S.~Cho, D.~Kim, K.~Kong, S.~H.~Lim, K.~T.~Matchev, J.~C.~Park and M.~Park,
 ``750 GeV Diphoton Excess May Not Imply a 750 GeV Resonance,''
  Phys.\ Rev.\ Lett.\  {\bf 116}, no. 15, 151805 (2016)
  [arXiv:1512.06824 [hep-ph]].
  
  \bibitem{Chao:2015nsm} 
  W.~Chao,
  ``Symmetries behind the 750 GeV diphoton excess,''
  Phys.\ Rev.\ D {\bf 93}, no. 11, 115013 (2016)
  [arXiv:1512.06297 [hep-ph]].
  
  \bibitem{Chakraborty:2015jvs} 
  I.~Chakraborty and A.~Kundu,
  ``Diphoton excess at 750 GeV: Singlet scalars confront triviality,''
  Phys.\ Rev.\ D {\bf 93}, no. 5, 055003 (2016)
  [arXiv:1512.06508 [hep-ph]].
  
\bibitem{Han:2015qqj} 
  X.~F.~Han and L.~Wang,
  ``Implication of the 750 GeV diphoton resonance on two-Higgs-doublet model and its extensions with Higgs field,''
  Phys.\ Rev.\ D {\bf 93}, no. 5, 055027 (2016)
  [arXiv:1512.06587 [hep-ph]].
  
  \bibitem{Bi:2015uqd} 
  X.~J.~Bi, Q.~F.~Xiang, P.~F.~Yin and Z.~H.~Yu,
  ``The 750 GeV diphoton excess at the LHC and dark matter constraints,''
  Nucl.\ Phys.\ B {\bf 909}, 43 (2016)
  [arXiv:1512.06787 [hep-ph]].
  
  \bibitem{Berthier:2015vbb} 
  L.~Berthier, J.~M.~Cline, W.~Shepherd and M.~Trott,
  ``Effective interpretations of a diphoton excess,''
  JHEP {\bf 1604}, 084 (2016)
  [arXiv:1512.06799 [hep-ph]].
  
  \bibitem{Bauer:2015boy} 
  M.~Bauer and M.~Neubert,
  ``Flavor anomalies, the 750 GeV diphoton excess, and a dark matter candidate,''
  Phys.\ Rev.\ D {\bf 93}, no. 11, 115030 (2016)
  [arXiv:1512.06828 [hep-ph]].
  
  \bibitem{Dev:2015isx} 
  P.~S.~B.~Dev and D.~Teresi,
  ``Asymmetric dark matter in the Sun and diphoton excess at the LHC,''
  Phys.\ Rev.\ D {\bf 94}, no. 2, 025001 (2016)
  [arXiv:1512.07243 [hep-ph]].
  
  \bibitem{Badziak:2015zez} 
  M.~Badziak,
  ``Interpreting the 750 GeV diphoton excess in minimal extensions of Two-Higgs-Doublet models,''
  Phys.\ Lett.\ B {\bf 759}, 464 (2016)
  [arXiv:1512.07497 [hep-ph]].
  
  \bibitem{Altmannshofer:2015xfo} 
  W.~Altmannshofer, J.~Galloway, S.~Gori, A.~L.~Kagan, A.~Martin and J.~Zupan,
  ``750 GeV diphoton excess,''
  Phys.\ Rev.\ D {\bf 93}, no. 9, 095015 (2016)
  [arXiv:1512.07616 [hep-ph]].
  
  \bibitem{Das:2015enc} 
  K.~Das and S.~K.~Rai,
  ``750 GeV diphoton excess in a U(1) hidden symmetry model,''
  Phys.\ Rev.\ D {\bf 93}, no. 9, 095007 (2016)
  [arXiv:1512.07789 [hep-ph]].
  
  \bibitem{Cao:2016cok} 
  Q.~H.~Cao, Y.~Q.~Gong, X.~Wang, B.~Yan and L.~L.~Yang,
  ``One bump or two peaks: The 750 GeV diphoton excess and dark matter with a complex mediator,''
  Phys.\ Rev.\ D {\bf 93}, no. 7, 075034 (2016)
  [arXiv:1601.06374 [hep-ph]].
  
  \bibitem{Harland-Lang:2016qjy} 
  L.~A.~Harland-Lang, V.~A.~Khoze and M.~G.~Ryskin,
  ``The production of a diphoton resonance via photon-photon fusion,''
  JHEP {\bf 1603}, 182 (2016)
  [arXiv:1601.07187 [hep-ph]].
  
  \bibitem{Aparicio:2016iwr} 
  L.~Aparicio, A.~Azatov, E.~Hardy and A.~Romanino,
  ``Diphotons from Diaxions,''
  JHEP {\bf 1605}, 077 (2016)
  [arXiv:1602.00949 [hep-ph]].
  
  \bibitem{Ko:2016wce} 
  P.~Ko and T.~Nomura,
  ``Dark sector shining through 750 GeV dark Higgs boson at the LHC,''
  Phys.\ Lett.\ B {\bf 758}, 205 (2016)
  [arXiv:1601.02490 [hep-ph]].
  
  \bibitem{D'Eramo:2016mgv} 
  F.~D'Eramo, J.~de Vries and P.~Panci,
  ``A 750 GeV Portal: LHC Phenomenology and Dark Matter Candidates,''
  JHEP {\bf 1605}, 089 (2016)
  [arXiv:1601.01571 [hep-ph]].
  
  \bibitem{Han:2016bus}
  X.~F.~Han, L.~Wang, L.~Wu, J.~M.~Yang and M.~Zhang,
  ``Explaining 750 GeV diphoton excess from top/bottom partner cascade decay in two-Higgs-doublet model extension,''
  Phys.\ Lett.\ B {\bf 756} (2016) 309
  [arXiv:1601.00534 [hep-ph]].
  
  \bibitem{Jiang:2015oms} 
  Y.~Jiang, Y.~Y.~Li and T.~Liu,
  ``750 GeV Resonance in the Gauged $U(1)'$-Extended MSSM,''
  Phys.\ Lett.\ B {\bf 759}, 354 (2016)
  [arXiv:1512.09127 [hep-ph]].
  
  \bibitem{Choi:2016cic} 
  S.~M.~Choi, Y.~J.~Kang and H.~M.~Lee,
  ``Diphoton resonance confronts dark matter,''
  JHEP {\bf 1607}, 030 (2016)
  [arXiv:1605.04804 [hep-ph]].
  
  \bibitem{Franceschini:2016gxv} 
  R.~Franceschini, G.~F.~Giudice, J.~F.~Kamenik, M.~McCullough, F.~Riva, A.~Strumia and R.~Torre,
  ``Digamma, what next?,''
  arXiv:1604.06446 [hep-ph].
  
  \bibitem{Badziak:2016cfd} 
  M.~Badziak, M.~Olechowski, S.~Pokorski and K.~Sakurai,
  ``Interpreting 750 GeV Diphoton Excess in Plain NMSSM,''
  Phys.\ Lett.\ B {\bf 760}, 228 (2016)
  [arXiv:1603.02203 [hep-ph]].
  
  \bibitem{Han:2016pab} 
  C.~Han, K.~Ichikawa, S.~Matsumoto, M.~M.~Nojiri and M.~Takeuchi,
  ``Heavy fermion bound states for diphoton excess at 750 GeV Ñ collider and cosmological constraints,''
  JHEP {\bf 1604}, 159 (2016)
  [arXiv:1602.08100 [hep-ph]].
  
  \bibitem{Hamada:2016vwk} 
  Y.~Hamada, H.~Kawai, K.~Kawana and K.~Tsumura,
  ``Models of the LHC diphoton excesses valid up to the Planck scale,''
  Phys.\ Rev.\ D {\bf 94}, no. 1, 014007 (2016)
  [arXiv:1602.04170 [hep-ph]].
  
\bibitem{DiLuzio:2016sbl} 
  L.~Di Luzio, F.~Mescia and E.~Nardi,
  ``Redefining the Axion Window,''
  Phys.\ Rev.\ Lett.\  {\bf 118}, no. 3, 031801 (2017)
  [arXiv:1610.07593 [hep-ph]].
  
  \bibitem{Han:2015cty} 
  C.~Han, H.~M.~Lee, M.~Park and V.~Sanz,
  ``The diphoton resonance as a gravity mediator of dark matter,''
  Phys.\ Lett.\ B {\bf 755}, 371 (2016)
  [arXiv:1512.06376 [hep-ph]].
  
\bibitem{PeskinSchroeder}
M.~Peskin and D.~Schroeder,
``An Introduction to Quantum Field Theory,"
Westview Press (1995).

\bibitem{Kang:2006yd} 
  J.~Kang, M.~A.~Luty and S.~Nasri,
  ``The Relic abundance of long-lived heavy colored particles,''
  JHEP {\bf 0809}, 086 (2008)
  [hep-ph/0611322].
  
\bibitem{KenLang}  
K.~Lang,
``The Sun from Space,"
Springer (2000).

\end{thebibliography}
\end{document}